\documentclass[
    reprint,
    superscriptaddress,
    amsmath,
    amssymb,
    aps,
    prx,
    nofootinbib
]{revtex4-2}

\usepackage{textcomp}  
\usepackage{graphicx}  
\usepackage{hyperref}  
\usepackage{siunitx}
\usepackage{xspace}
\usepackage{tabularx}
\usepackage[symbol]{footmisc}

\newcommand{\detectionFidelity}{$99.971(1)\%$}
\newcommand{\imagingSurvival}{$99.80(5)\%$}
\newcommand{\ax}{$a_x=579(2)\,$nm}
\newcommand{\ay}{$a_y=1187(18)\,$nm}
\newcommand{\vacuumLifetime}{$273(3)\,$s}
\newcommand{\fullLatticeDepth}{$1.47\,$mK}

\makeatletter
\def\maketitle{
\@author@finish
\title@column\titleblock@produce
\suppressfloats[t]}
\makeatother

\begin{document}

\newcommand{\partitle}[1]{\section{#1}}

\newcommand{\papertitle}{High-fidelity detection of large-scale atom arrays in an optical lattice}

\title{\papertitle{}}

\author{Renhao Tao}\thanks{These authors contribute equally to this work.}
    \affiliation{Max-Planck-Institut f\"{u}r Quantenoptik, 85748 Garching, Germany}
    \affiliation{Munich Center for Quantum Science and Technology (MCQST), 80799 Munich, Germany}
  \affiliation{Fakultät für Physik, Ludwig-Maximilians-Universit\"{a}t, 80799 Munich, Germany}
  
\author{Maximilian Ammenwerth}\thanks{These authors contribute equally to this work.}
    \affiliation{Max-Planck-Institut f\"{u}r Quantenoptik, 85748 Garching, Germany}
    \affiliation{Munich Center for Quantum Science and Technology (MCQST), 80799 Munich, Germany}

\author{Flavien Gyger}\thanks{These authors contribute equally to this work.}
    \affiliation{Max-Planck-Institut f\"{u}r Quantenoptik, 85748 Garching, Germany}
    \affiliation{Munich Center for Quantum Science and Technology (MCQST), 80799 Munich, Germany}

\author{\\Immanuel Bloch}
    \affiliation{Max-Planck-Institut f\"{u}r Quantenoptik, 85748 Garching, Germany}
    \affiliation{Munich Center for Quantum Science and Technology (MCQST), 80799 Munich, Germany}
    \affiliation{Fakultät für Physik, Ludwig-Maximilians-Universit\"{a}t, 80799 Munich, Germany}

\author{Johannes Zeiher}
\email{johannes.zeiher@mpq.mpg.de}
    \affiliation{Max-Planck-Institut f\"{u}r Quantenoptik, 85748 Garching, Germany}
    \affiliation{Munich Center for Quantum Science and Technology (MCQST), 80799 Munich, Germany}
    \affiliation{Fakultät für Physik, Ludwig-Maximilians-Universit\"{a}t, 80799 Munich, Germany}

\date{\today}

\begin{abstract}
Recent advances in quantum simulation based on neutral atoms have largely benefited from high-resolution, single-atom sensitive imaging techniques.
A variety of approaches have been developed to achieve such local detection of atoms in optical lattices or optical tweezers.
For alkaline-earth and alkaline-earth-like atoms, the presence of narrow optical transitions opens up the possibility of performing novel types of Sisyphus cooling, where the cooling mechanism originates from
the capability to spatially resolve the differential optical level shifts in the trap potential.
Up to now, it has been an open question whether high-fidelity imaging could be achieved in a ``repulsive Sisyphus" configuration, where the trap depth of the ground state exceeds that of the excited state involved in cooling.
Here, we demonstrate high-fidelity (\detectionFidelity) and high-survival (\imagingSurvival) imaging of strontium atoms using repulsive Sisyphus cooling.
We use an optical lattice as a pinning potential for atoms in a large-scale tweezer array with up to $399$ tweezers and show repeated, high-fidelity lattice-tweezer-lattice transfers.
We furthermore demonstrate loading the lattice with approximately 10000 atoms directly from the MOT and scalable imaging over $>10000$ lattice sites with a combined survival probability and classification fidelity better than 99.2\%.
Our lattice thus serves as a locally addressable and sortable reservoir for continuous refilling of optical tweezer arrays in the future. 

\end{abstract}

\maketitle


Laser-cooled atomic gases trapped in optical lattices have enabled a number of breakthroughs in quantum sciences~\cite{Gross2017, Morgado2021, Katori2011}.
An entirely new level of control of such systems was reached by the development of quantum-gas microscopes~\cite{Bakr2009, Sherson2010}.
%
These setups have enabled single-site- and single-atom-resolved detection of atomic many-body systems in a top-down approach starting from a quantum-degenerate gas prepared via evaporative cooling.
%
A prerequisite to quantum-gas microscopy is the high-fidelity and low-loss imaging of atoms in optical lattices.
For alkali atoms, cooling during imaging can be achieved either by polarization gradient cooling~\cite{Bakr2009, Sherson2010} or Raman sideband cooling~\cite{Haller2015,Cheuk2015,Omran2015}.
%
In alkaline-earth and alkaline-earth-like atoms, the presence of a narrow optical intercombination transition opens up the perspective for new, efficient cooling strategies~\cite{Norcia2018,Cooper2018,Covey2019,Saskin2019, Urech2022}.
In particular, high-resolution imaging of atoms in optical lattices has been achieved for ytterbium atoms~\cite{Yamamoto2016},
and recently for strontium atoms in a clock-magic optical lattice at $813\,$nm~\cite{Schine2022,Young2022,Young2023}. 

Atom assembly in arrays of optical tweezers provides an alternative, bottom-up approach for the study of many-body systems with single-atom preparation, control and detection capabilities~\cite{Schlosser2001, Grunzweig2010,Kaufman2021}.
This approach benefits from the re-configurable design of array patterns in various dimensions~\cite{Endres2016,Barredo2016,Barredo2018}, as well as the ability of single-site addressing and atom positioning~\cite{Barredo2016,Barredo2018,Bluvstein2022}.
These features have resulted in successful implementations of tweezer arrays in various fields such as quantum metrology~\cite{Madjarov2019, Norcia2019, Young2020}, quantum computing~\cite{Levine2019,Bluvstein2022,Graham2022,Jenkins2022,Ma2022,Ma2023,Evered2023} and quantum simulation~\cite{Scholl2021,Ebadi2021}.    
While bottom-up and top-down approaches have been developed mostly in parallel, increasing efforts have recently been undertaken to combine both platforms, leading to novel ways of preparing atoms in optical lattices in the Hubbard regime~\cite{Trisnadi2022, Young2022, Young2023}, coupling freely configurable optical tweezer arrays for realizing Hubbard models~\cite{Murmann2015,Spar2022} or creating novel programmable optical lattice potentials via selective blocking of specific sites in optical lattices~\cite{Wei2023}.
This hybrid approach has also played a role in scaling neutral-atom systems by allowing optimal use of different potentials for distinct experimental stages, for instance, in creating programmable arrays of $\sim100$ optical qubits in magic wavelength potentials \cite{Schine2022, Eckner2023}.

\begin{figure*}
    \centering
    \includegraphics[width=\textwidth]{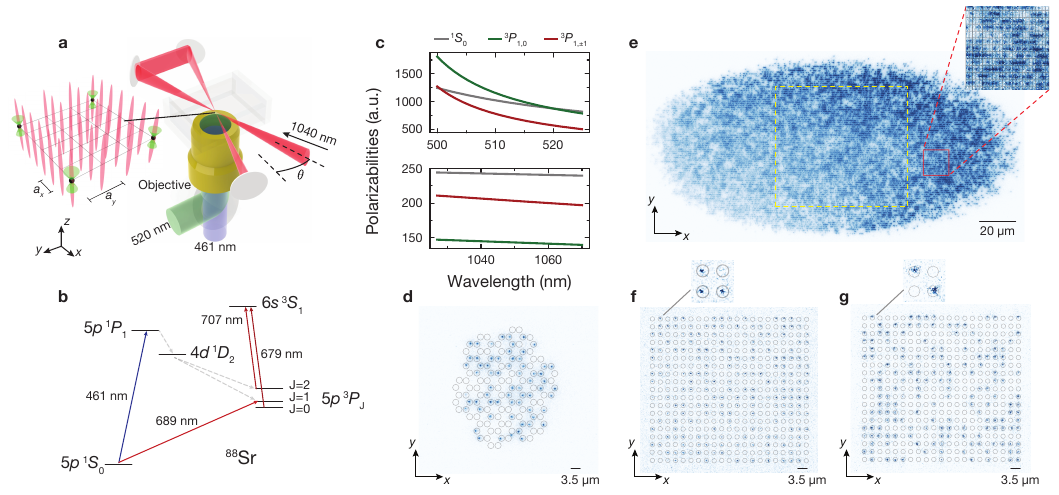}
    \caption{
    \textbf{Experimental setup.} \textbf{a} An optical lattice is formed from a single elliptical beam at wavelength $1040\,$nm retro-reflected in the $xy$ plane. At the crossing angle $\theta = 26^\circ$, the lattice spacing is \ax\  along the $x$-axis and \ay\  along the $y$-axis. The tweezer array (green) is created with a spatial light modulator (SLM) and light at wavelength $520\,$nm. The tweezers are focused into the glass cell and overlapped with lattice sites (red) in 3D (inset). The fluorescence of atoms is collected by the objective and separated from the tweezer light with a dichroic. 
    \textbf{b} Level diagram for transitions involved in this work. We resonantly scatter photons on the $^1$S$_0$-$^1$P$_1$ transition ($461\,$nm), while simultaneously cooling on the $^1$S$_0$-$^3$P$_1$ transition ($689\,$nm). Atoms leaking out of the imaging cycle are repumped via transitions at $679\,$nm and $707\,$nm. 
    \textbf{c} Calculated polarizabilities of $^1$S$_0$ and $^3$P$_{1, m_J}$ at green tweezer wavelengths (upper panel) and infrared lattice wavelengths (lower panel). At trapping wavelengths ranging from $500\,$ to $520\,$nm,  $^1$S$_0$-$^3$P$_{1, m_J=0}$ can be made magic via mixing of the polarizabilities for different Zeeman sublevels under strong magnetic field~\cite{Norcia2018}. The mixing ratio can be tuned via the angle $\phi$ between the bias field and the linear tweezer polarization. At $1040\,$nm, cooling on the narrow line occurs in the repulsive Sisyphus regime.
    \textbf{d} Typical single-shot image of atoms loaded from a honeycomb-shaped tweezer array and imaged in the lattice. Circles denote the programmed tweezer location. 
    \textbf{e} A single-shot image of more than $10000$ single atoms directly loaded into the lattice from the magneto-optical trap. 
    The yellow dashed box denotes the spatial extent of the tweezer array used in this work. The smaller red box is a zoom-in which shows well-resolved single atoms. 
    \textbf{f} Typical single-shot image of a $21 \times 19$ tweezer array imaged directly in $520\,$nm tweeezer. The tweezer spacing is $3.478(1)(3.549(1))\,\mu$m along the $x(y)$-axis which is precisely chosen to be $6a_x(3a_y)$, respectively. 
    \textbf{g} Typical single-shot image of atoms loaded from a tweezer array into the lattice and imaged there. Due to weaker axial ($z$ axis) confinement,  the point-spread function of atoms imaged in the lattice is about $1.7$ times larger than that in tweezers.
    }
    \label{fig:1}
\end{figure*}
%
%
Here, we demonstrate preparation and detection of $10^4$ single atoms using a hybrid lattice-tweezer platform.
This significant advance in system size for the field of atomic arrays is enabled by several innovations going beyond previous work.
%
%
In particular, employing a specially purposed trapping geometry and a previously unexplored lattice trapping wavelength for strontium atoms at $1040\,$nm, we demonstrate high-fidelity, low-loss detection of the atoms in the optical lattice.
We quantify the imaging performance in this novel configuration and report a classification fidelity exceeding $99.9\%$ combined with a survival probability exceeding $99.29(1)\%$ averaged over $10450$ sites of the lattice.
Our work surpasses the state of the art in demonstrating the largest number of traps amenable to high-fidelity and low-loss imaging to date~\cite{Ebadi2021, Young2023, Lars2023}.
We furthermore show that a dense cloud of atoms can be directly loaded into a single plane of the lattice from a magneto-optical trap, and subsequently imaged with high fidelity and low loss, opening the path to an entirely different approach to scaling atom arrays. 
In addition, we demonstrate repeated handover between the lattice and tweezers generated using a spatial light modulator (SLM), and reinitialization of the tweezer-trapped atoms in low-temperature states after the imaging step in the lattice.
Finally, our results settle an ongoing discussion raised by earlier work~\cite{Taieb1994,Cooper2018,Holzl2023} on whether repulsive Sisyphus cooling and high-fidelity and low-loss imaging are compatible.\\  
%
In our experiment, we combine an optical tweezer array comprising $399$ optical traps at wavelength $\lambda_{tw} = 520\,$nm and an optical lattice operated at wavelength $\lambda_l = 1040\,$nm, see Fig.~\ref{fig:1}a.
Using computer-generated holograms, we routinely create tweezer arrays with trap spacings of $3.478(1)(3.549(1))\,\mu$m along the $x(y)$-axis and waists of $473(3)\,$nm, with excellent control over positioning, spacing and arrangement of the individual traps, see~\cite{SI}.
The optical lattice is formed in a bow-tie configuration~\cite{Sebby-Strabley2006}, where a single beam creates a two-dimensional lattice potential by four-fold interference.
In addition, we tightly focus the lattice in the $z$-axis to a waist of $20\,\mu$m, which provides a vertical confinement of up to $5.9\,$kHz at a lattice depth of \fullLatticeDepth.
The half-angle $\theta$ is about $26^\circ$ chosen to have lattice constants ratio of 1:2 along two axes.
In this configuration, the radial trap frequencies are $150\,$kHz and $300\,$kHz respectively, such that we achieve complete 3D confinement using a single lattice beam only.
Cooling in the various configurations described above is performed with a single beam addressing the $^1$S$_0$-${}^3$P$_1$ transition at $689\,$nm, see Fig.~\ref{fig:1}b.
For imaging, we additionally illuminate the atoms with light at $461\,$nm, which induces fluorescence on the broad $^1$S$_0$-$^1$P$_1$ transition.
We collect the fluorescence photons with the same objective that is used to generate the tweezer array with a specified NA$=0.65$. This has allowed us to resolve atoms spaced as closely as one lattice site \ax, see Fig.~\ref{fig:1}.\\
\begin{figure}
    \centering
    \includegraphics{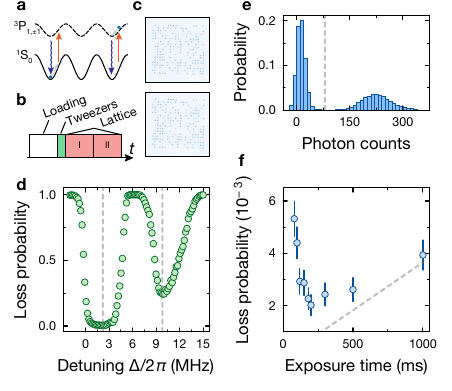}
    \caption{
    \textbf{Imaging in the lattice.} 
    \textbf{a} Illustration of repulsive Sisyphus cooling dynamics for $^1$S$_0$ and $^3$P$_1$ states in the lattice. 
    \textbf{b}. Schematic of experimental sequence for imaging in the lattice. Two images (I \& II) in the lattice are taken after atoms are prepared in tweezers and transferred into the lattice. Loss probability is calculated by comparing occupations between image I and II.
    \textbf{c}. Exemplary consecutive fluorescence images of individual atoms in the lattice showing no loss. \textbf{d}. Array-averaged atom loss probability as a function of $689\,$nm cooling laser detuning $\Delta$ from the $^1$S$_0$-$^3$P$_1$ resonance in free space for an imaging exposure time of $300\,$ms. 
    At $\Delta/2\pi=2.2\,$MHz, imaging loss reduces down to $2\times10^{-3}$. 
    \textbf{e} Array-averaged histogram of photon counts taken with exposure time 300 ms, showing a well-resolved background and one-atom fluorescence peak. 
    The classification fidelity can be as high as \detectionFidelity, see~\cite{SI}.
    \textbf{f}. Imaging loss at constant scattered photon number for classification infidelity of $10^{-4}$ vs. exposure time. The lower dashed gray line indicates atom loss probability attributed to our estimated vacuum lifetime of \vacuumLifetime, which is reached by our imaging in the limit of long exposure time and low illumination power. The background level remains similarly low all all exposure times.}
    \label{fig:2}
\end{figure}
As a first step, we demonstrate high-fidelity and low-loss imaging in a repulsive Sisyphus regime.
This is relevant for strontium at our lattice wavelength of $1040\,$nm, where $\omega_g/\omega_e=1.08$ for the $^3$P$_{1, m_J = \pm 1}$ as the excited state, see Fig.~\ref{fig:2}a.
We begin the experiment with an array of atoms at $399$ singly-occupied lattice sites loaded from tweezers, see~\cite{SI}.
To optimize the cooling performance for imaging, we scan the detuning $\Delta$ of the cooling light relative to the free-space resonance, see Fig.~\ref{fig:2}d.
We obtain a broad cooling feature at approximately $2.2\,$MHz, where the atom loss fraction reaches the sub-percent, and a second narrower feature at about $9.8\,$MHz, where the loss is higher.
The two features can be attributed to cooling on $^1$S$_0$-$^3$P$_{1, m_J= \pm 1}$ and $^1$S$_0$-$^3$P$_{1, m_J=0}$ respectively and are consistent with the 689 nm transition split by tensor lightshift in the $1040\,$nm lattice.
To characterize imaging performance under cooling on the $m_J = \pm1$ transition, we take two consecutive images in the optical lattice, see Fig.~\ref{fig:2}c.
The images are binarized based on the tweezer-averaged histogram of the integrated photon count, see Fig.~\ref{fig:2}e.
With an optimal threshold~\cite{SI}, we obtain a classification infidelity of approximately $10^{-4}$, demonstrating the feasibility of high-fidelity imaging in our lattice.
To benchmark the atom loss probability from imaging, we compare the occupation of the two consecutively acquired images as a function of the exposure time, see Fig.~\ref{fig:2}f.
For this measurement, we keep the integrated photon number scattered on the $461\,$nm transition and hence the classification fidelity constant.
We find a robust minimum atom loss probability at an exposure time of approximately $200\,$ms, where the loss reaches $2\times10^{-3}$.
At shorter exposure times and hence larger imaging beam scattering rate, the atom loss probability increases as a result of recoil heating exceeding the cooling rate from Sisyphus cooling.
For longer exposure times, the atom loss probability begins to be dominated by our estimated vacuum lifetime of \vacuumLifetime.\\
%
The feasibility of high-fidelity, low-loss imaging in the lattice offers the perspective of decoupling the power-intensive imaging step from cooling and physics performed in optical tweezers, provided efficient transfer between lattice and tweezer array.
Such a capability would allow for the use of advantageous features of tweezers at $520\,$nm, e.g.\ for trapping of Rydberg states via the ionic core polarizability~\cite{Wilson2022}, while avoiding lossy detection at the same wavelength~\cite{Cooper2018, Holzl2023}.
We characterize the lattice-tweezer transfer via the round-trip atom loss probability after imaging first in the lattice, see Fig.~\ref{fig:3}a.
A challenge in this case is the weak vertical confinement of our 2D lattice, whose waist in the $z$-direction significantly exceeds the Rayleigh range of $1.5\,\mu$m of the tweezers, see Fig.~\ref{fig:3}a.
To enable low-loss transfer back to tweezers, we first perform an optimized repulsive Sisyphus cooling in the lattice after imaging (II).
Subsequently, we ramp up the tweezers to a depth of $300\,\mu$K, before lowering the lattice to an intermediate depth of $150\,\mu$K.
We perform a second stage of cooling in this combined potential to efficiently transfer the atoms into trapped states in the tweezers.
The cooling frequency is chosen to coincide with the lattice-light shifted cooling sideband of the tweezers and the magnetic field is set to the magic cooling transition in tweezers alone~\cite{SI}.
Finally, we ramp the lattices down in $50\,$ms, completing the transfer to the tweezers.
Imaging is then performed once more in the lattice, with an identical tweezer-lattice handover as before the first image.
We benchmark the complete round-trip atom loss probability $p_n$ by comparing the reconstructed tweezer occupation between two images taken in the lattice before and after $n$ transfers, see Fig.~\ref{fig:3}c.
While the overall atom loss probability increases with the number of round-trips as expected, we find that the atom loss probability per round-trip $p_1$, extracted under the assumption of a simple power-law scaling of the atom loss probability $1-p_n = (1-p_1)^n$, continuously decreases from $1.3\%$ down to approximately $5\times10^{-3}$ after a few round-trips.
We attribute the initially higher atom loss probability predominantly to a systematic spatial inhomogeneity of the lattice potential affecting our cooling in the lattice during transfer, which becomes directly apparent in a tweezer-resolved transfer loss map after $n=80$ round-trips, see Fig.~\ref{fig:3}b inset.
Hence, we consider the reported transfer loss as a worst case scenario that can be improved by excluding the traps exhibiting high atom loss or centering the tweezer array in the lattice.
We find that highly efficient transfers are possible if the tweezer depth in the transfer exceeds approximately $300\,\mu$K, see Fig.~\ref{fig:3}c.
For the last point beyond $400\,\mu$K, the transfer loss probability increases slightly due to non-optimal cooling parameters. 
\begin{figure}
    \centering
    \includegraphics{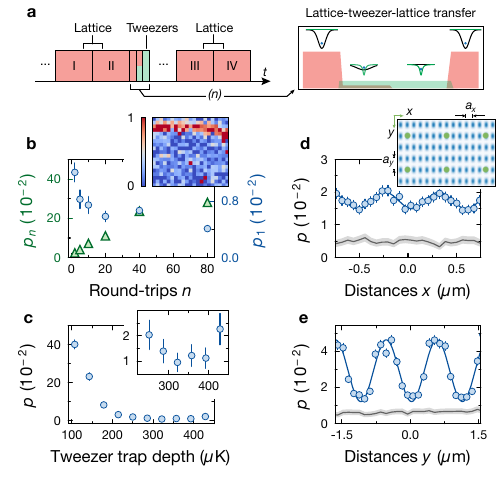}
    \caption{\textbf{Transfer between lattice and tweezers.} 
    \textbf{a} Schematic of experimental sequence for tweezer-lattice-tweezer transfers. Axial potential landscape of the tweezers (green) and lattice (black) during hand-over is shown in the inset. The sketech is not to scale.
    \textbf{b} Cumulative loss probability $p_n$ (green triangles) of atoms versus the number of lattice-tweezers-lattice round-trips $n$. The average single round-trip atom loss probability $p_1$ (blue round markers) decreases as $n$ increases. The atom loss probability $p_n$ shows a pronounced spatial dependence predominantly at the boundary of the lattice, as apparent from a measurement of the site-resolved atom loss probability after $n=80$ round-trips (inset). 
    \textbf{c} Single-round-trip atom loss probability $p$ as function of tweezer trap depth after transfer. The inset shows a close-up and confirms tweezer averaged single-round-trip losses close to $1\%$. 
    \textbf{d,e} Single-round-trip atom loss probability $p$ vs. relative position between lattice and tweezer potentials along $x$-axis (\textbf{d}) and $y$-axis (\textbf{e}) shown as blue points. The sinusoidal fit reflects the expected lattice potential with a lattice constant \ax\ and \ay. The atom loss probability due to imaging alone is indicated by shaded gray lines. Inset: Sketch of the tweezer traps (green dots) and the lattice potential (blue dots).
     }
    \label{fig:3}
\end{figure}
To study the dependence of the transfer efficiency on the relative position between lattice sites and tweezers, we scan the position of the tweezer array along either the $x$- or $y$-axis, see Fig.~\ref{fig:3}d, e.
We find a pronounced sinusoidal dependence of the transfer loss, which reaches up to $5\%$ in non-optimal conditions.
Close to the optimal condition,  we obtain a $\sim 90\%$ fidelity of finding an atom at exactly the same lattice sites before (II) and after (III) holding them in the tweezers.
The observed sinusoidal structure is in excellent agreement with the expected dependence for our lattice, and the curves represent a characterization of the underlying lattice structure using a large-scale tweezer array~\cite{Deist2022}. 
We note that even after imaging in the lattice, one can re-cool atoms in tweezers close to the radial motional ground state after transferring them back~\cite{SI}.\\
%
\begin{figure}
\includegraphics{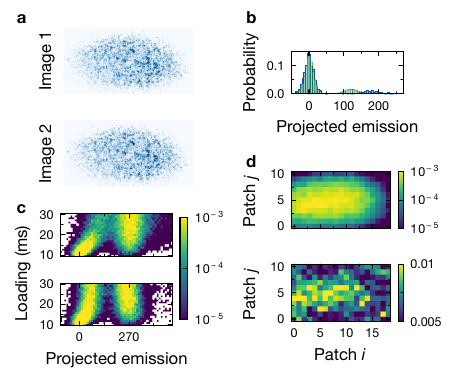}
    \caption{
    \textbf{Imaging characterization over the entire lattice loaded directly from the MOT}
 \textbf{a} 
 We load the lattice directly from the MOT and take two consecutive images to characterize the imaging performance over a much larger region of the lattice. 
 The sparse filling gives rise to a higher classification fidelity (see discussion below) and aids in the bench-marking.
 \textbf{b} Coarse-grained histograms of two representative lattice patches ($10\times5$ sites) show the spread of the emissions due to fluorescence inhomogeneity.
 \textbf{c}
The detected counts versus MOT loading time for two consecutive images reveals the parity projection for an exposure time of $1.8\,$s, signalled by the absence of the tail of the histograms extending into high emission counts at large loading times in the second image. 
The double-occupancy accounts for less than $0.3\%$ of all emissions events at $30\,$ms loading time. 
%
%
\textbf{d}
 Coarse-grained classification infidelity (upper panel) analyzed on individual patches in the entire lattice. The classification error is below $10^{-3}$. 
 The higher infidelity at the lattice center correlates well to the atom density, as a result of cross-talk that worsens at higher lattice filling fraction.
The coarse-grained loss probability (lower panel) of two consecutive images demonstrates that the imaging loss computed from consecutive images is as low as $7.1(1)\times 10^{-3}$. 
To further optimize the performance, the exposure time was set to $900\,$ms for this measurement. 
    }
    \label{fig:4}
\end{figure}
To highlight the scalability of our platform, we directly load the lattice from the MOT for variable durations and characterize the imaging performance for the loaded atoms over two consecutive images, see Fig~\ref{fig:4}a. 
Importantly, our lattice configuration renders further single-plane preparation steps before detecting the atomic distribution unnecessary.
To benchmark the detection, we binarize the images using deconvolution techniques, which also yields the classification fidelity, see~\cite{SI}.
We evaluate the histogram of recorded photon counts locally, over small patches of $10\times5$ sites to mitigate the effect of inhomogeneities, see Fig~\ref{fig:4}b.
With increasing loading time, a single-atom peak in the histogram of the first acquired image develops at $270$ photon counts, corresponding to the amount of photons collected within the exposure time for each singly-loaded site, see Fig~\ref{fig:4}c. 
At even longer loading time exceeding $30\,$ms, a tail beyond $270$ counts appears to extend into higher emission counts, which is expected for loading multiple atoms per site. 
In the second image, this tail is absent even at long loading times, indicating efficient parity projection during the imaging process.
To quantify the achieved performance, we characterize the classification fidelity and survival probability in a region of $10450$ lattice sites. 
The coarse-grained analysis indicates that the high lattice filling and consequently cross-talk between sites is the main factor that reduces classification fidelity at long loading time, see Fig~\ref{fig:4}d. 
In particular, the higher classification infidelity can be attributed to an increasing width of the zero-atom peak due to the empty sites receiving fluorescence emanating from adjacent occupied sites. 
Nevertheless, we find that classification infidelity is globally below $10^{-3}$ for a filling fraction $\sim 0.2$ which amounts to $\sim 2300$ loaded atoms. 
Similarly, the imaging loss is kept at $7.1(1)\times 10^{-3}$, see Fig~\ref{fig:4}d.
%
%
%
%
The preparation and high-fidelity detection of individual atoms in a single layer of an optical lattice allows for subsequent resorting in the lattice as demonstrated recently in the same setup~\cite{Gyger2024}.
Such sorted arrays can then be transferred into tweezer potentials, resulting in a direct two-fold reduction of the required tweezer power due to then deterministic loading of the tweezer array with near unity-filling. 
Using additional vertical confinement in the transfer allows for significantly relaxed power requirements of the tweezer array, promising further gains in the scalability of tweezer arrays through deterministic loading via optical lattices.

In conclusion, we have demonstrated the feasibility of low-loss and high-fidelity imaging under repulsive Sisyphus cooling conditions on the narrow-linewidth transition of strontium in an optical lattice.
We extend the size of the system compatible with single-site and single atom detection to more than 10000 lattice sites and load more than 10000 atoms directly from the MOT.
Our results offer a new path to assembling large atom arrays in optical lattices that clearly surpass the state of the art with respect to the achieved atom numbers in sortable optical tweezers and lattices~\cite{Ebadi2021, Young2023, Lars2023}.
Straightforward upgrades of the laser power used in our setup via commercially available off-the-shelf laser systems allow to scale the number of sites by a factor of $10$, as a direct consequence of our proof-of-concept demonstration of high imaging quality at $1040\,$nm, where such laser systems are readily available.
Furthermore, our work offers the perspective to operate tweezer arrays at arbitrary wavelengths by decoupling the power-intensive imaging step from preparation and physics in optical tweezer arrays, with potential applications in quantum simulation of Ising models~\cite{Scholl2021,Ebadi2021,Slagle2022, Scholl2023}, lattice gauge theories~\cite{Homeier2023}, quantum chemistry~\cite{Malz2023}, or quantum-enhanced 
metrology~\cite{Schine2022,Eckner2023,Bornet2023,YoungJ2023} in scalable ensembles.
Finally, directly loading the optical lattice from a magneto-optical trap in combination with high-fidelity imaging, resorting and laser-cooling, provides a new bottom-up approach of assembling large-scale Hubbard simulators~\cite{Young2022,Young2023}.

\begin{acknowledgments}
We thank Sylvain {de L{\'e}s{\'e}leuc} and Stepan Snigirev for insightful discussions on tweezer arrays, Isabella Fritsche and the planqc team for help in setting up a high-power tweezer laser setup, Elias Trapp for support in the lab, and David Wei for providing the code for lattice reconstruction.
We acknowledge funding by the Max Planck Society (MPG) the Deutsche Forschungsgemeinschaft (DFG, German Research Foundation) under Germany's Excellence Strategy--EXC-2111--390814868, and from the Munich Quantum Valley initiative as part of the High-Tech Agenda Plus of the Bavarian State Government.
This publication has also received funding under Horizon Europe programme HORIZON-CL4-2022-QUANTUM-02-SGA via the project 101113690 (PASQuanS2.1).
J.Z. acknowledges support from the BMBF through the program “Quantum technologies - from basic research to market” (Grant No. 13N16265).
F.G. acknowledges funding from the Swiss National Fonds (Fund Nr P500PT\textunderscore203162). M.A. and R.T. acknowledge funding from the International Max Planck Research School (IMPRS) for Quantum Science and Technology. M.A acknowledges support through a fellowship from the Hector Fellow Academy.

\end{acknowledgments}

\appendix

\setcounter{figure}{0}
\renewcommand\theequation{S\arabic{equation}}
 \renewcommand\thefigure{S\arabic{figure}}

\section{Supplementary Information}
\subsection{Details on the Experimental Setup}
\begin{figure}[h]
    \centering
\includegraphics{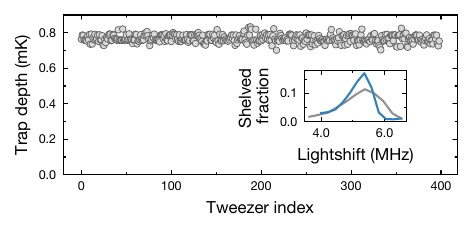}
    \caption{
    \textbf{Tweezer-resolved trap depth after feedback.} With three rounds of equalization using lightshift measurements, the root-mean-square trap depth inhomogeneity reaches about $2.2\%$ for $399$ tweezers.  The array-averaged $^1$S$_0$-$^3$P$_{1,|m_J|=1}$ shelving spectra before (gray) and after corrections (blue) are shown in the inset. The linewidth narrowing shows a smaller spread of trap depths as a result of the homogenization.}
    \label{fig:s:tweezer_inhomogeneity}
\end{figure}

The optical lattice in our experiment is created with a single $1040\,$nm beam retro-reflected in 4f-configuration on the $xy$-plane, see Fig.~1a. 
The beam crossing angle $\theta$ is chosen to be $26^\circ$ resulting in a lattice spacing \ax\ along the $x$-axis and \ay\ along the $y$-axis.
The foci of all four beams overlap at the center of the field of view of the objective with waists of $20\,\mu$m ($100\,\mu$m) along the $z$-axis (in the $xy$-plane). 
With the polarization of all beams aligned along $z$-axis, multiple-pass interference yields a trap frequency that can reach $150\,$kHz ($300\,$kHz) along the $x$-axis ($y$-axis), at a trap depth of $1.47\,$mK with $12\,$W optical power. 
In the $z$-axis, the harmonic confinement of the lattice reaches about $2\pi \times 5.9\,$kHz.
The tweezer spacing in both axes is precisely chosen to be $6a_y$ and $3a_x$, respectively. 
Using $650\,$mW optical power at 520 nm, we create a 2D array of $21 \times 19= 399$ tweezers that have trap depth $140\,\mu$K each. 
The waist of the tweezers is estimated to be $472(3)\,$nm from independent calibrations of trap depth and trap frequency.
The trap frequency at 140~$\mu$K tweezers depth is about $\omega_r/2\pi = 77.6(5)\,$kHz radially as measured from sideband spectra. From tweezer waist and radial trap frequency, we deduce an axial trap frequency of $\omega_z /2\pi= 19.2(5)\,$kHz.  
The trap depths for $399$ tweezers are equalized to a root-mean-square intensity variation to $2.2\%$ by minimizing variations in measured differential lightshifts on the ${}^1$S$_0$-${}^3$P$_1$ transition of an initial tweezer pattern pre-corrected for diffraction efficiency, see Fig~\ref{fig:s:tweezer_inhomogeneity}.
We carefully superimpose the tweezer array with the lattice by aligning their foci in all three dimensions. 
The global tweezer array position on the $xy$-plane is adjusted actively by feedback to a diffraction grating phase on the spatial light modulator to maintain optimal overlap with the slowly drifting lattice potential.  

To initialize experiments involving tweezers, we first load a $140\,\mu$K-deep tweezer array with atoms from a single-frequency magneto-optical trap~\cite{Norcia2018b} on the $689\,$nm transition, followed by the parity projection in tweezers. 
The lattice is on during the loading state for loading enhancement.
The filling fraction is typically $45\%$.
Next, atoms are cooled to the radial ground state in the tweezers via sideband cooling in a magic magnetic field of $14\,$G oriented at an angle $\phi = 5.6^\circ$ relative to the linear tweezer polarization along the $y$-axis~\cite{Yamamoto2016}.
This leaves the atoms in the ground state along the radial directions with an estimated average ground-state population of $97^{+2}_{-5} \%$.
For imaging characterization in the lattice, we then transfer the atoms from the tweezers to the optical lattice by ramping up the lattice trap depth and subsequently ramping down the tweezer trap depth in $5\,$ms.

\subsection{Tweezer hologram generation}
\label{sec:twe-hol}

\subsubsection{Tweezer generation}
We generate our tweezer array holographically using a phase-only spatial-light modulator.
The algorithm for creating the hologram largely follows earlier work~\cite{Kim2019} and involves a variant of the Gerchberg–Saxton algorithm where the phases of tweezers in the focal plane are kept stationary towards the end of the optimization. 
In the optical setup, we use a collimated beam with a waist of $4.2\,$mm incident on the SLM chip of dimension $12.8\,$mm$\times15.9\,$mm. 
A telescope consisting of a $300\,$mm and $750\,$mm achromatic lens relays the imprinted phase pattern to the back focal plane of the high-resolution objective. 
The objective then transforms the phase pattern into an array of tightly-focused tweezers in the atomic plane. 

\begin{figure*}
    \centering
\includegraphics{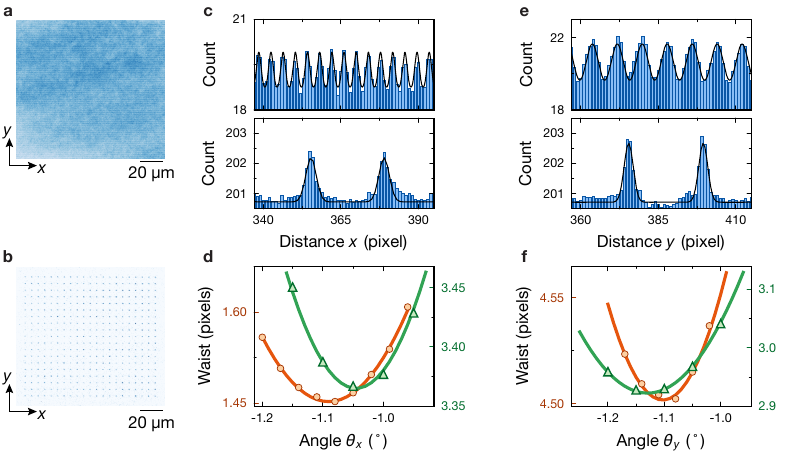}
    \caption{\textbf{Determination of spacings and coordinate angles of tweezers and lattice.} \textbf{a} An averaged fluorescence image of a sparsely-loaded lattice used for the analysis. \textbf{b} An averaged fluorescence image of a tweezer array imaged in tweezers. For images of 2D fluorescence patterns defined by two orthogonal symmetry axes (like \textbf{a} and \textbf{b}), the 1D fluorescence count, which results from projecting the image array along some axis, exhibits smaller peak overlap when the projection axis coincides with either symmetry axis of the array~\cite{Sherson2010,Yamamoto2016}. Such a case for $x$($y$)-axis is shown for lattice in the upper panel and the tweezer in the lower panel in \textbf{c}(\textbf{e}) Any deviation from such a condition results in an increased peak overlap (\textbf{d}, \textbf{f}). For the analysis, the dependency of waist on the angle of projection axis is fitted with $\sigma_0\left[1+\beta\{ 1-\cos(\gamma (\theta_i-\theta_{i, 0})) \} \right]$ for $i=x, y$~\cite{Yamamoto2016}. In both \textbf{d} and \textbf{f}, lattice data (circle) and fit (solid line) are red, whereas tweezer data (triangles) and fit (solid line) are green. Once the minimum-waist angle is found, the lattice and tweezer array spacing and waist can be extracted by fitting a grid of Gaussians functions in the rotated frame of reference (\textbf{c}, \textbf{e}). In summary, we extract lattice coordinate angles $\theta_x^\text{lat}(\theta_y^\text{lat}) = -1.090(0)^\circ (-1.102(1)^\circ)$ and tweezer coordinate angles $\theta^\text{tw}_x (\theta^\text{tw}_y) = -1.0452(2) (-1.1296(1))^\circ$. For the spacings, we find lattice constants of $a_x(a_y) = 3.92(6) (8.04(9))$ pixels and tweezers spacing of $c^\text{tw}_x(c^\text{tw}_y) = 23.564(6)(24.047(6))$ pixels. }
    \label{fig:s:tweezer_lattice_spacing_angles}
\end{figure*}
Undiffracted light at the SLM is blocked in an intermediate imaging plane.
An added blazing phase grating steers the diffracted beam by about $0.5^\circ$ away from the undiffracted light and defines the new optical axis, with which we align our optics downstream. 
To avoid clipping at the input aperture of the objective, we impose a circular aperture of $11.5\,$mm in diameter on the SLM, such that the light fills less than $90\,\%$ of the input pupil of the objective~\cite{Salter2020}. 
This aperture is assumed in all phase-retrieval algorithms. 
To correct wavefront distortion, we sample the SLM light shortly before it enters the objective and focus the beam onto a camera.
Using phase-shifting interferometry \cite{Zupancic2016,Tsevas2021} with super-pixels on the SLM, we are able to suppress the RMS wavefront error to below $\lambda/14$.  

\subsubsection{Tweezer amplitude equalization}
\label{sec:twe-tw_homogenization}
The first step to equalizing the individual depths of the optical tweezers consists of a compensation of the finite diffraction efficiency of the SLM, which can be understood as a consequence of the diffraction effect on the finite-sized SLM pixels. 
The diffraction efficiency $\eta(\theta_x, \theta_y)$ can be calibrated as a function of diffraction angle $(\theta_x,\theta_y)$ using a diagnostic camera. 
The target tweezer amplitude is then weighted by $1/\sqrt{\eta}$ in the phase-retrieval algorithm~\cite{Kim2019}. 
As a second step, we use the atomic signal to spectroscopically probe the homogeneity of the tweezer array via the differential light shift induced on the cooling transition.
At our tweezer wavelength of $520\,$nm, the $\sigma_\pm$-component of the $689\,$nm transition exhibits about $7.3\,\mathrm{MHz}/\mathrm{mK}$ differential lightshift, such that it is sensitive to spatial variations in tweezer depth.  
The spectroscopy is performed at a magnetic bias field of $14\,$G along the tweezer polarization axis to break the Zeeman degeneracy in the ${}^3$P$_1$ state. 
We adapt our optimization procedure for the two-step equalization protocol.
The first step consists of $40$ iterations of Gerchberg–Saxton algorithm (GSA), five iterations of weighted GSA, and another $35$ iterations where the phase of each tweezer on the image plane is frozen.
Tweezer arrays created at this point can typically reach less than about $8\%$ intensity inhomogeneity for our $399$ tweezer array.
Subsequently, we spectroscopically probe the tweezer depth $U_i$, where $i$ denotes the tweezer index, and adjust each tweezer amplitude by a factor $\overline{\sqrt{U_i}}/\sqrt{U_i}$, where $\overline{\sqrt{U_i}}$ is the value averaged over all $i$. 
After this second round of homogenization, the tweezer intensity inhomogeneity can typically reach below $2.2\%$ after three iterations, see Fig~\ref{fig:s:tweezer_inhomogeneity}. 

\subsubsection{Determining spacings and angles of tweezers and lattice potentials}
\label{sec:spacing_angles}

To create a tweezer array commensurate with the lattice, we need to determine the lattice coordinate system.
For this, we largely follow the work of \cite{Sherson2010, Yamamoto2016}, and base all analysis on an averaged fluorescence images of a sparsely-populated lattice loaded from the MOT, see Fig~\ref{fig:s:tweezer_lattice_spacing_angles}a. 
We eliminate a small mismatch in the rotation angles of the lattice and tweezers relative to the camera pixel grid by rotating the image, followed by a 1D averaging along one of the two directions.
The alignment is optimal when the 1D averaged images display the highest contrast (or smallest fringe-width). 
Then, by fitting a grid of Gaussian profiles to the fluorescence peaks, we can identify the spacing. 
Such an analysis is done for both axes independently, yielding lattice spacings $a_x(a_y) = 3.92(6) (8.04(9))$ pixels and coordinate angles $\theta_x^\text{lat}(\theta_y^\text{lat}) = -1.090(0)^\circ (-1.102(1)^\circ)$, see Fig.~\ref{fig:s:tweezer_lattice_spacing_angles}.  From the known wavelength of the lattice laser of $1040.139(1)\,$nm in vacuum, we can deduce that one camera pixel is equal to $147.6(23)\,$nm in the atomic plane. The lattice constants expressed in SI are then 
\ax\ and \ay, respectively.

Once the coordinate system of the lattice is known, it is fairly straightforward to shape the tweezer array such that both potentials are commensurate. 
We define tweezer spacing $c^\text{tw}_i$ such that $c^\text{tw}_x = 6 a_x$ and $c^\text{tw}_y = 3 a_y$, and $\theta_i^\text{tw} = \theta_i^\text{lat}$ for both $x$ and $y$. 
With fluorescence imaging in tweezers only, we are able to verify the ratio of spacings directly.
With a similar analysis as for the lattice, we find that the tweezer array has spacing  $c^\text{tw}_x(c^\text{tw}_y) = 23.564(6)(24.047(6))$ pixels respectively and coordinate angles $\theta^\text{tw}_x (\theta^\text{tw}_y) = -1.0452(2) (-1.1296(1))^\circ$, see Fig~\ref{fig:s:tweezer_lattice_spacing_angles}. 

\subsection{Radial ground-state re-cooling after imaging in the lattice}
Our tweezers have well-resolved radial sidebands and are therefore conducive to ground-state cooling. After atoms are imaged in the lattice and transferred back to tweezers, they in general acquire an elevated temperature, as indicated by the prominent cooling sideband in the sideband thermometry, see Fig~\ref{fig:SI_4}a. We show that we can still perform resolved sideband cooling on the radial degree of freedom and obtain atoms with $98_{-3}^{+0}\%$ ground-state fraction. This is comparable to the fraction we can achieve in tweezers by itself without involving the lattice at all. We cross-check the cooling with the release-and-recapture in tweezers, see Fig~\ref{fig:SI_4}b. The result is consistent with the temperature estimated from sideband thermometry.
\begin{figure}
    \includegraphics{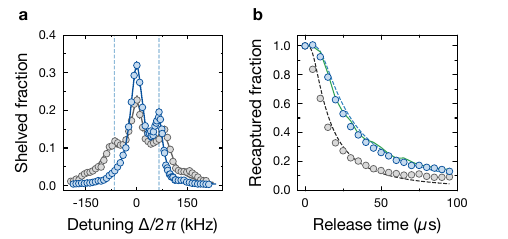}
    \caption{
    \textbf{Reinitialization in radial motional ground-state after transfer.} \textbf{a} Radial sideband spectra of atoms loaded back into tweezers without (gray) and with (blue) sideband cooling after transfer. From the asymmetry of the spectrum, we extract ground-state fraction $98_{-3}^{+0}\%$ with and $37_{-17}^{+40}\%$ without sideband cooling after the transfer.
    \textbf{b} Release-and-recapture measurements in tweezers. We record the recovered fraction of atoms after a variable release time with (blue) and without (gray) sideband cooling after transfer. From a 3D Monte Carlo simulation, we extract temperatures of $5\,\mu$K ($17\,\mu$K) with (without) sideband cooling after transfer, which is consistent with the temperature deduced from mean motional excitation measured by sideband spectroscopy. The fits are shown as blue and gray dashed lines respectively. The result of a release-and-recapture measurement before transferring to the lattice is indicated in green. 
    }
    \label{fig:SI_3}
\end{figure}
\subsection{Laser cooling on narrow optical transitions}
Narrow optical transitions found in alkaline-earth or alkaline-earth-like atomic species are well suited for a variety of cooling techniques that leverage differential trapping between the two states coupled via the narrow transition~\cite{Taieb1994,Cooper2018,Covey2019,Jackson2020,Urech2022,Holzl2023}.
Generally, three conditions can be distinguished: (i) the trap frequency of the excited state $\omega_e$ is larger than that of the ground state, $\omega_g$ (``attractive Sisyphus regime", $\omega_e>\omega_g$), (ii) the trap frequency of the excited state is smaller than that of the ground state (``repulsive Sisyphus regime", $\omega_e<\omega_g$) and (iii) both the excited and ground state are equally trapped (``magic regime", $\omega_e=\omega_g$).
In the magic trapping regime, sideband cooling can be performed, where motional quanta are removed in a controlled way from the atoms by the light tuned to the resonance of the cooling sideband at a detuning $\Delta=-\omega_e$ from the free-space resonance at $\Delta=0$.
Here, equal and strongly confining trapping potentials for both ground and excited states are essential as this allows atoms to change their internal states without significant motional heating.
However, even for unequal polarizabilities, one can find closed cycles of atoms undergoing excitation and deexcitation, during which the net kinetic energy can be reduced and converted to energy removed by emitted light.
For attractive Sisyphus cooling, the cooling light is tuned to the \textit{red} of the free-space resonance ($\Delta<0$), and optimized to resonantly excite atoms residing at the trap center, whereas for repulsive Sisyphus cooling, the light is tuned to the \textit{blue} of the free-space resonance ($\Delta>0$), and optimized to resonantly excite atoms away from the trap center.
Note that in the case of repulsive Sisyphus cooling, the light is still detuned to the red of the light-shifted resonance in the center of the trap, setting an effective ``Sisyphus cap" in the detuning, above which an atom would be heated and repelled from the trap~\cite{Cooper2018}.
In consequence, whereas attractive Sisyphus cooling features an intrinsically stable cooling configuration and provides cooling independent of atom temperature, repulsive Sisyphus cooling can lead to runaway heating in case the temperature and hence potential energy of an atom exceeds the Sisyphus cap, see Fig~1e. 

\subsection{Estimating classification fidelity from histogram fitting}
\begin{figure}
    \includegraphics{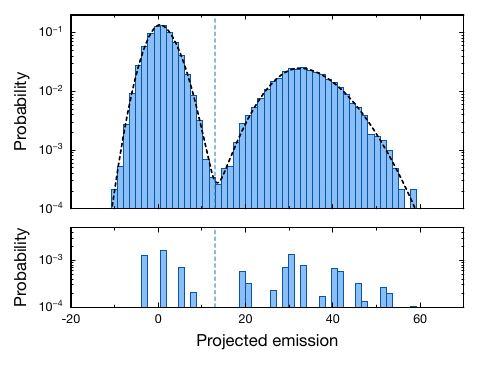}
    \caption{
    \textbf{Classification fidelity estimate from histogram fitting} 
    The fluorescence of individual atoms is de-convolved with a measured point-spread function and summed over a region of interest spanning $3\times3$ lattice sites. The resultant counts (in units proportional to photons) from $60$ shots are then aggregated in a histogram (upper panel) on logarithmic scale to reveal low probability events. 
    We fit the histogram with two Gaussian-like functions (black, dashed) and find a threshold (blue, dashed) that simultaneously minimize the false positive and false negative rate. The classification fidelity is then estimated to be \detectionFidelity. The fitting residual (lower panel) is small and indicates the accuracy of the model.
    }
    \label{fig:SI_4}
\end{figure}
To measure the survival probability in imaging with low error, we need an accurate protocol of distinguishing the presence of an atom from the background. 
A common technique involves summing up fluorescence counts scattered over a region of interest (roi) centered at individual atoms and comparing the result against a threshold~\cite{Cooper2018, Urech2022}. 
To increase the signal-to-noise ratio in the detection, we first of all apply a de-convolution filter derived from the measured point-spread function to individual atomic fluorescence before summing up the counts in a square roi $3\times3$ lattice sites large. 
On account of our use of tweezers in loading the lattice, the position of the atoms in the lattice is susceptible to small variation from shot to shot, as a result of relative time-dependent phase drift and imperfect time-independent incommensurability between the two potentials. 
We heuristically find that identifying the location of atoms with a lattice sites-resolved occupation reconstruction aids in centering the roi. 
Consequently, the histogram exhibits a smooth profile that is amenable to analytical modeling, see Fig~\ref{fig:SI_4}. 
We empirically fit the one-atom emission and background separately using a modified Gaussian function $P(x, x_0) = A \exp\left[ -\frac{(x-x_0)^n}{\sigma^2}\right]$ where $n$ is allowed to deviate from 2 and differs depending on whether $x<x_0$ or $x \geq x_0.$ 
An optimal threshold can be found by minimizing the probability of detecting false positive and false negative. 
We find that the final classification fidelity is insensitive to the histogram bin size. 
The error bar in the classification is estimated from assigning the fitting residual to falsely detected events according to the worse case scenario.

\nocite{Norcia2018b,Sherson2010, Salter2020, Zupancic2016,Tsevas2021, Yamamoto2016,
Taieb1994,Covey2019,Jackson2020,Urech2022,Holzl2023,Cooper2018, Urech2022, Kim2019}

\bibliography{LatticeEnhancedTweezers}

\end{document}